\documentclass[10pt,conference]{IEEEtran}
\usepackage{cite}
\usepackage{amsmath,amssymb,amsfonts}
\usepackage{algorithmic}
\usepackage{graphicx}
\usepackage{textcomp}

\usepackage{color}
\usepackage{xspace}
\newcommand{\systemname}{\textsc{CHASE}}
\newcommand{\doublecirc}{\ooalign{$\bigcirc$\crcr\hss$\circ$\hss}}

\usepackage{enumitem}
\setlist[itemize]{leftmargin=*,itemsep=0pt,parsep=0pt}

\usepackage{caption}
\usepackage{listings}
\usepackage{xcolor}
\definecolor{codecomment}{rgb}{0,0.5,0}
\definecolor{codekeyword}{rgb}{0.58,0,0.82}
\definecolor{codestring}{rgb}{0.58,0,0}
\definecolor{deobfuscator-orange}{RGB}{241,158,57}
\definecolor{webresearcher-blue}{RGB}{89,133,225}
\usepackage{wasysym}
\usepackage{multirow}
\usepackage{tabularx,ragged2e,siunitx}
\usepackage{fontawesome5}
\usepackage{mathtools}
\usepackage[hidelinks,colorlinks=true,linkcolor=blue,citecolor=blue]{hyperref}
\hypersetup{
    colorlinks=true,
    linkcolor=blue,
    citecolor=blue,
    filecolor=blue,
    urlcolor=blue,
}

\begin{document}

\title{CHASE: LLM Agents for\\Dissecting Malicious PyPI Packages}

\author{
  \IEEEauthorblockN{Takaaki Toda}
  \IEEEauthorblockA{\textit{Department of Computer Science and Engineering} \\
    \textit{Waseda University}\\
    Tokyo, Japan \\
    todatakaaki@nsl.cs.waseda.ac.jp}
  \IEEEauthorblockA{\textit{RIKEN AIP}\\
    Tokyo, Japan}
  \and
  \IEEEauthorblockN{Tatsuya Mori}
  \IEEEauthorblockA{\textit{Department of Computer Science and Engineering} \\
    \textit{Waseda University}\\
    Tokyo, Japan \\
    mori@nsl.cs.waseda.ac.jp}
  \IEEEauthorblockA{\textit{RIKEN AIP}\\
    Tokyo, Japan}
  \IEEEauthorblockA{\textit{NICT}\\
    Tokyo, Japan}
}

\maketitle

\begin{abstract}
  Modern software package registries like PyPI have become critical infrastructure for software development, but are increasingly exploited by threat actors distributing malicious packages with sophisticated multi-stage attack chains. While Large Language Models (LLMs) offer promising capabilities for automated code analysis, their application to security-critical malware detection faces fundamental challenges, including hallucination and context confusion, which can lead to missed detections or false alarms. We present CHASE (Collaborative Hierarchical Agents for Security Exploration), a high-reliability multi-agent architecture that addresses these limitations through a Plan-and-Execute coordination model, specialized Worker Agents focused on specific analysis aspects, and integration with deterministic security tools for critical operations. Our key insight is that reliability in LLM-based security analysis emerges not from improving individual model capabilities but from architecting systems that compensate for LLM weaknesses while leveraging their semantic understanding strengths.
  Evaluation on a dataset of 3,000 packages (500 malicious, 2,500 benign) demonstrates that CHASE achieves 98.4\% recall with only 0.08\% false positive rate, while maintaining a practical median analysis time of 4.5 minutes per package, making it suitable for operational deployment in automated package screening.
  Furthermore, we conducted a survey with cybersecurity professionals to evaluate the generated analysis reports, identifying their key strengths and areas for improvement.
  This work provides a blueprint for building reliable AI-powered security tools that can scale with the growing complexity of modern software supply chains.
  Our project page is available at: \url{https://t0d4.github.io/CHASE-AIware25/}
\end{abstract}


\section{Introduction}

Software package registries have revolutionized modern software development by enabling developers to freely share and reuse code~\cite{zimmermann2019small}. The Python Package Index (PyPI), serving millions of developers worldwide, exemplifies this collaborative ecosystem~\cite{kikas2017structure}. However, this openness has been increasingly exploited by threat actors who distribute malicious packages containing malware~\cite{ohm2020backstabber, taylor2020defending}. The sheer scale of modern software supply chains---with millions of packages and billions of downloads---makes manual security analysis impractical. Despite efforts by PyPI's security team to establish reporting mechanisms, automated detection remains a significant challenge, with repository administrators requiring exceptionally low false positive rates~\cite{vu2023bad}.

Current malware detection approaches primarily employ binary classification (benign vs. malicious) but struggle with sophisticated modern malware that uses multi-stage attack chains, encrypted payloads, and dynamic code obfuscation~\cite{sejfia2022practical, duan2023towards}. Conventional static analysis cannot handle dynamic code construction, while dynamic analysis faces resource constraints and evasion techniques, necessitating automated solutions that match modern attack complexity. While Large Language Models (LLMs) show promise for automated security analysis through their code semantic understanding~\cite{fang2024}, their application presents significant SE4AI challenges. LLMs suffer from hallucinations and context confusion that can result in missed detections or false alarms, undermining the reliability required for production deployment~\cite{jiang2024, he2025}. Single LLM agents often fail to maintain a consistent analysis state across complex, multi-stage malware behaviors.

To address these challenges, we propose \systemname{} (\underline{C}ollaborative \underline{H}ierarchical \underline{A}gents for \underline{S}ecurity \underline{E}xploration), a high-reliability multi-agent architecture that combines Plan-and-Execute workflow, specialized Worker Agents, and robust deterministic tools. Our approach mitigates LLM limitations through architectural design: a supervisor agent maintains global analysis state and prevents context confusion, specialized worker agents focus on specific analysis aspects to reduce hallucination, and deterministic tools provide ground truth for critical operations like deobfuscation and URL fetching. The Plan-and-Execute pattern enables adaptive analysis strategies that can handle the dynamic nature of modern malware.

Our key insight is that reliability in LLM-based security analysis comes not from improving individual model capabilities, but from architecting systems that compensate for LLM weaknesses while leveraging their strengths. By combining multiple specialized agents with deterministic tools and structured coordination patterns, we achieve both the semantic understanding needed for complex malware analysis and the reliability required for practical deployment.

This paper makes the following contributions:
\begin{itemize}
    \item We identify and characterize the fundamental challenges of applying autonomous LLM agents to security analysis tasks, demonstrating how hallucination and context confusion specifically impact malware detection reliability.
    
    \item We design and implement \systemname{}, a multi-agent architecture that achieves high reliability through Plan-and-Execute coordination, specialized Worker Agents, and integration with deterministic security tools.
    
    \item We demonstrate \systemname{}'s operational effectiveness by achieving 98.4\% recall with only 0.08\% false positive rate using a dataset of 3,000 real-world packages, while maintaining a practical median analysis time of 4.5 minutes per package suitable for automated screening. Retrying on timeouted cases further improved recall.

    \item We evaluate the generated analysis reports through a survey of cybersecurity professionals, identifying their key strengths and areas for improvement.
\end{itemize}

Our work demonstrates that well-architected multi-agent systems can overcome the limitations of individual LLMs in security analysis. As software development increasingly incorporates AI assistance, our approach provides a blueprint for building reliable AI-powered security tools that can scale with the growing complexity of software supply chains.

\section{Background}
\label{sec:background}

\subsection{The PyPI Ecosystem and the Malicious Package Threat}
As Python has become increasingly prevalent in machine learning and web development, the number of packages on PyPI, its primary package registry, has grown dramatically to nearly 640,000 hosted packages~\cite{pypi_stats_dashboard}. PyPI's ease of use, allowing anyone to upload packages and providing pip as a convenient package manager, has accelerated modern software development. Additionally, various package managers with different features such as Pipenv, Poetry, Rye, and uv have emerged, further enhancing PyPI's utility.

However, this accessibility has been exploited by malicious actors. In March 2024, PyPI was forced to temporarily suspend new package uploads after a mass upload of typosquatting malware targeting popular packages~\cite{pypi_infostealer_influx}. More alarmingly, in November 2024, an AWS credential-stealing package was discovered to have existed on PyPI since March 2021, accumulating 37,100 downloads~\cite{hackernews_fabrice}. Even legitimate packages are not immune: in December 2024, the official ultralytics package for the renowned YOLO object detector was compromised with cryptominer code through a GitHub Actions vulnerability~\cite{pypi_ultralytics_attack}.

The threat landscape is further complicated by packages that transition from benign to malicious through updates~\cite{semantic_types_malicious}, and PyPI's package name transfer policy, as defined in PEP 541~\cite{pep541}, which allows unmaintained packages to be taken over. Despite PyPI's efforts through the Malware Reporting and Response Project~\cite{pypi_malware_reporting_evolved} and Project Quarantine~\cite{pypi_project_quarantine}, automation remains a critical challenge. Prior research surveying package manager operators~\cite{vu2023bad} identified high false positive rates as a major barrier to introducing automatic scanners.

\subsection{Anatomy of a Malicious Package}
As an example of a sophisticated modern malicious package, Listing~\ref{listing:malpkg-ethereim-example} shows a portion of the \verb|setup.py| file from \texttt{ethereim-1.0.0}, formerly available on PyPI.

\begin{lstlisting}[language=Python, caption={Malicious package example (ethereim-1.0.0). }, label=listing:malpkg-ethereim-example]
class ynSmToSfOIO...cgJswETrR(install):
  def run(self):
    import os
    if os.name == "nt":
      import requests
      from fernet import Fernet
      exec(
        Fernet(b'e2_3wb...8Adk_rY=')
        .decrypt(b'gAAAA...7OZwvw=')
      )
    install.run(self)
\end{lstlisting}

During installation, the run function checks for a Windows environment (line 4), then uses the Fernet library to decrypt and execute a hidden payload (lines 7-10). The resulting decrypted code (Listing~\ref{listing:ethereim-decrypted-payload}) reveals the attacker's true intent:

\begin{lstlisting}[language=Python, caption={Decrypted payload showing the attacker's true intent. Schema modified to hxxps for safety.}, label=listing:ethereim-decrypted-payload]
exec(
  requests.get(
    "hxxps://funcaptcha.ru/paste2?package=ethereim"
  ).text.replace("<pre>", "").replace("</pre>", "")
)
\end{lstlisting}

This multi-stage pattern, where the package initially acts as a downloader for subsequent remote payloads, demonstrates why automated analysis must consider dynamic code construction and execution flows.

\subsection{Current Detection Approaches and Their Limitations}

Existing research on malicious package detection falls into three categories, each with inherent limitations:

\noindent\textbf{Static Analysis Approaches.} These methods analyze code structure without execution. Sun et al.~\cite{SunXiaobing20241IDC} combine BERT-processed call graphs and control flow graphs with metadata. Zhang et al.~\cite{ZhangJunan2025KTBw} leverage cross-language similarities between Python and JavaScript malware, converting function call sequences to natural language descriptions for BERT processing. While computationally efficient, static analysis cannot handle dynamic code construction and is increasingly vulnerable to plausible LLM-generated metadata.

\noindent\textbf{Dynamic Analysis Integration.} Fewer approaches incorporate runtime analysis. Zheng et al.~\cite{ZhengXinyi2024TRDo} execute packages in Docker containers to capture system and API call traces for rule-based matching. Huang et al.~\cite{HuangCheng2024DmNp} combine static pre-filtering with dynamic verification for suspicious packages. While resistant to obfuscation, dynamic analysis faces resource constraints and sophisticated evasion techniques like time-delayed payloads.

\noindent\textbf{LLM-Enhanced Detection.} Recent works apply LLM to malware detection. Zahan et al.~\cite{ZahanNusrat2025LLLM} demonstrate GPT-4's effectiveness using Chain-of-Thought prompting on JavaScript packages. Others use LLMs for feature extraction: Huang et al.\cite{HuangYiheng2024SPDo} for identifying sensitive APIs, Wang et al.~\cite{WangJian2025MALM} for automatic feature generation, and Gao et al.~\cite{GaoXingan2025USENIX} for high-risk API identification through graph centrality analysis.
Despite these advances, current approaches fall short of revealing complete attack chains, ultimate malicious objectives, and Indicators of Compromise (IoCs). They typically identify suspicious code fragments without deobfuscating code or retrieving remote payloads to understand the attackers' final intentions.

\subsection{Agentic AI for Complex Analysis} %
LLM agents represent a paradigm shift from single-turn interactions to autonomous, multi-step problem solving. Initially experimental, by early 2025, they had matured into practical tools. Unlike standalone LLMs, agents possess two critical capabilities: decomposing complex problems into executable steps and interacting with the real world through tools.
Multi-agent architectures further enhance these capabilities, enabling collaborative problem-solving at scale, as demonstrated by commercial Deep Research services~\cite{openai_deep_research}. In software development, tools like Codex~\cite{openai_codex} showcase agents' ability to understand entire codebases and perform sophisticated analysis.
This evolution positions AI agents as ideal candidates for malicious package analysis, where understanding requires navigating multi-stage payloads, deobfuscating code, retrieving remote components, and synthesizing findings into actionable intelligence. These tasks align perfectly with agents' autonomous, tool-using capabilities.

\section{\systemname{} Architecture}

\subsection{Overall Design Philosophy}

\begin{figure}[tbp]
    \centering
    \includegraphics[width=\columnwidth]{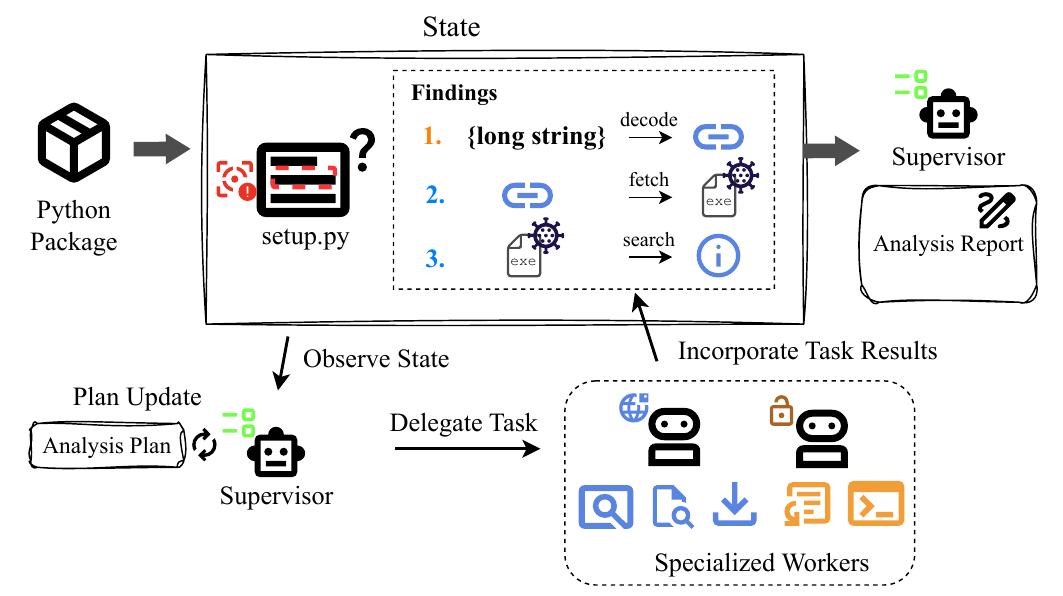}
    \caption{The multi-agent architecture of \systemname{}, using a Plan-and-Execute workflow and a Supervisor-Worker model.}
    \label{fig:system-architecture}
\end{figure}

Figure~\ref{fig:system-architecture} illustrates the architecture of \systemname{}, our autonomous agent system for malware analysis. Our design philosophy stems from understanding how human security experts analyze malicious code. Prior research~\cite{Lebiere2015functional, Thomson2015Malware} reveals that professional analysts dynamically switch between two cognitive modes: exploratory activities (surveying code to form hypotheses) and exploitative activities (focusing on specific tasks like deobfuscation).

This cognitive process is emulated by two design patterns:

\noindent\textbf{Supervisor-Worker Multi-Agent Architecture.} A single Supervisor agent maintains the global analysis perspective and orchestrates the investigation, while specialized Worker agents execute domain-specific tasks. This separation mirrors the macro-micro perspective switching of human analysts.

\noindent\textbf{Plan-and-Execute Workflow.} Following Wang et al.~\cite{wang-etal-2023-plan}, our Supervisor dynamically adjusts its analysis plan based on Worker results. This is crucial for malware analysis, where outcomes are unpredictable—deobfuscating code or fetching remote payloads often reveals unexpected paths requiring strategy adjustment.

\subsection{System Components}

\subsubsection{The Supervisor Agent}

The Supervisor orchestrates the entire analysis by:
\begin{itemize}
    \item Generating high-level task descriptions for Workers (specifying \emph{what} to analyze, not \emph{how})
    \item Integrating Worker responses into a coherent understanding
    \item Dynamically adjusting the plan based on discoveries
    \item Producing the final analysis report
\end{itemize}

This abstraction allows the Supervisor to focus on strategic planning without being encumbered by the  implementation details of specific analysis techniques.

\subsubsection{Worker Agents and Their Specializations}

Each Worker agent specializes in a specific analysis domain:

\noindent\textbf{Deobfuscator:} Handles code deobfuscation, decryption, and safe execution of suspicious code segments.

\noindent\textbf{Web Researcher:} Investigates external resources, fetches remote payloads, and queries threat intelligence databases.

Workers receive role specifications through system prompts, ensuring they understand their responsibilities and boundaries, in line with the principle of separation of concerns.

\subsection{Reliability-Oriented Agent Coordination}

To ensure reliable analysis, our architecture is specifically designed around core coordination mechanisms that directly mitigate known LLM weaknesses, such as hallucination and context confusion.

\subsubsection{State-Based Communication for Context Integrity}

Conventional multi-turn conversations that accumulate message history are susceptible to context bloat and the ``Lost in the Middle'' problem~\cite{LiuNelsonF.2024LitM}. This approach also risks context poisoning, where a single hallucination can contaminate all subsequent interactions~\cite{comanici2025gemini25pushingfrontier}. To mitigate these issues, \systemname{} avoids passing entire message lists and instead maintains all information in a central state variable. The multi-agent architecture isolates each agent's context, localizing any hallucination effects, while the Supervisor's review of Worker outputs provides additional error detection opportunities.

This state-driven approach significantly improves system efficiency and focus. After each Worker’s action, the system updates the state and generates a focused prompt containing only essential context for the next task, which decreases token consumption and helps Workers maintain better task focus. Variables in the state include source codes, a log of executed sub-tasks, and pending analysis tasks, ensuring no critical findings are lost. Furthermore, this enhances system robustness by enabling graceful recovery from Worker failures; the Supervisor can regenerate a task prompt from the preserved state without losing the entire conversation context, a crucial feature when analyzing heavily obfuscated source codes.

\subsubsection{Specialized Agents and Adaptive Planning to Prevent Failures}

LLMs, especially local ones, suffer performance degradation when given too many tools~\cite{ParamanayakamVaratheepan2025LiMO} or very long contexts~\cite{hong2025context}. Maintaining minimal, focused toolsets for each specialized Worker ensures stable performance.

Moreover, unlike reactive workflows like ReAct~\cite{YaoShunyu2023RSRA} that get trapped in repetitive failures (e.g., trying to decode random strings as base64), our Plan-and-Execute approach recovers via global goal awareness and dynamic replanning. To prevent the infinite loops common in multi-agent systems~\cite{pan2025why} without hard iteration limits that risk incomplete analyses, we use budget-aware planning. The Supervisor tracks an iteration budget, prioritizing high-value tasks as it depletes. For instance, if deobfuscation reveals a suspicious URL, the Supervisor pivots from string analysis to new network investigation tasks, such as checking its reputation. This adaptive replanning is crucial for uncovering multi-stage attacks and escaping local failures.

\section{Implementation}

\subsection{Model Selection}

LLM agents consume significantly more tokens than standard conversational AI. Anthropic~\cite{anthropic_multiagent_research} reports 15× higher consumption for multi-agent systems.
Given the high token consumption and the need to scalably analyze the vast number of packages uploaded daily to PyPI, proprietary models are economically infeasible. We therefore employ a carefully selected hierarchy of local LLMs via SGLang~\cite{sglang_official} to balance utility with cost.
For the Supervisor agent, we selected \textbf{Qwen3:32B} due to its superior multi-step reasoning capabilities and large 32K token context window, essential for maintaining coherent analysis state across complex investigations. The model's strong instruction-following ability enables it to effectively orchestrate multiple Worker agents while dynamically adjusting plans based on emerging findings. In contrast, Worker agents use the smaller \textbf{Qwen3:8B}, which provides sufficient capability for focused, domain-specific tasks while offering approximately 3× faster inference speed (tokens/s). This speed advantage is crucial for iterative operations like deobfuscation attempts. For structured output conversion, we employ \textbf{Gemma3:4B}, which excels at reliably converting free-form analysis results into JSON format with minimal latency.

This tiered approach effectively reduces total execution time compared to using a single large model for all roles, while maintaining comparable analysis quality on a single GPU.

\subsection{Minimal Toolset Design}

Following research showing that excessive tools degrade LLM performance~\cite{ParamanayakamVaratheepan2025LiMO}, we implement only essential tools, each designed for specific attack patterns commonly observed in malicious packages discovered on PyPI.

\noindent{\bf Deobfuscator Tools.}
The \texttt{decrypt\_fernet\_payload} tool handles Fernet symmetric encryption, found in approximately 40\% of malicious packages, with a detailed error message for LLM when decryption fails. The \texttt{decode\_base64\_payload} tool handles base64-encoded payloads—a common evasion technique used in the initial stage. For more sophisticated obfuscation techniques, such as those like JJEncode~\cite{Ferrie2011}, which utilize complex arithmetic operations, the \texttt{execute\_python\_code} tool safely executes code fragments in a sandboxed environment with an execution time limit of 1 minute.

\begin{figure}[tbp]
    \centering
    \includegraphics[width=0.8\columnwidth]{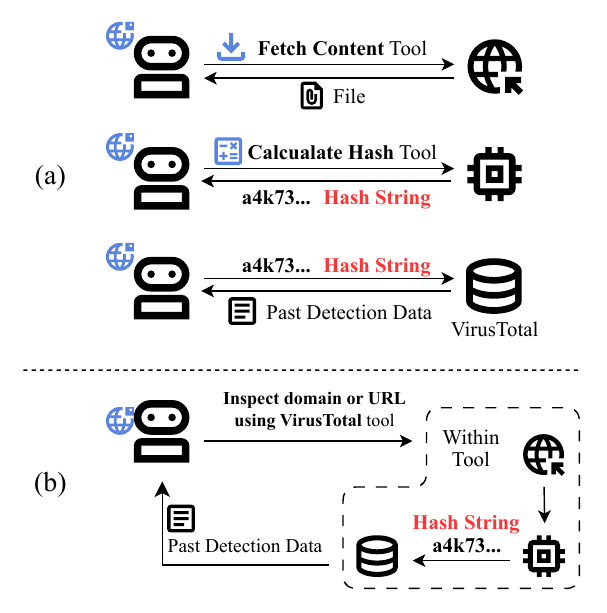}
    \caption{Design Principle for Reliable LLM-Tool Interaction: (a) A fine-grained tool design forces the LLM to handle high-entropy strings (e.g., hash values, shown in red), which is a hallucination-prone task. (b) Our composite tool, \texttt{inspect\_domain\_or\_url\_using\_virustotal}, encapsulates this entire workflow, shielding the LLM from volatile data handling to enhance overall system reliability.}
    \label{fig:tool-design-philosophy}
\end{figure}

\noindent{\bf Web Researcher Tools.}
The \texttt{TavilySearch} tool enables identification of known malware campaigns and threat actor patterns through web searches, with relevance filtering to prevent information overload. Since most sophisticated PyPI malicious packages use staged payloads, the \texttt{fetch\_content\_at\_url} tool retrieves remote content with size limits and format detection capabilities. The \texttt{inspect\_domain\_or\_url\_using\_virustotal} tool provides authoritative threat intelligence data about a given URL or domain name, and if given a URL pointing to a remote file, internally handles hash calculations to enhance reliability as illustrated in Figure~\ref{fig:tool-design-philosophy}.

\section{Evaluation}

\subsection{Experimental Setup}

\noindent
\textbf{Dataset.}
The dataset used to evaluate our agent's performance consists of 3,000 packages: 500 malicious and 2,500 benign, reflecting the minority of malicious packages in the real world. The malicious packages were collected from pypi\_malregistry created in~\cite{GuoWenbo2023AESo}, which is publicly available on GitHub at \url{https://github.com/lxyeternal/pypi_malregistry} and has been continuously updated since the publication of~\cite{GuoWenbo2023AESo}. We selected 500 packages from the latest commit at the time of the experiment (commit hash starting with {\tt 153aba56}). The 2,500 benign packages were randomly selected from newly published packages on PyPI from July to September in 2025. We confirmed that all packages in the performance evaluation dataset contain both \texttt{setup.py} and \texttt{\_\_init\_\_.py} files.

\noindent
\textbf{Analysis Configurations.}
We limited the scope of our analysis to \texttt{setup.py}, \texttt{\_\_init\_\_.py}, and Python source files imported by \texttt{\_\_init\_\_.py} for two primary reasons. First, these files represent the most common malicious entry points, as a prior study \cite{GuoWenbo2023AESo} has shown that 87.6\% of malicious packages execute payloads upon installation (via \texttt{setup.py}) or import (via \texttt{\_\_init\_\_.py}). Second, analyzing all files within a package was computationally prohibitive for the current implementation of \systemname{}.
We experimented on a server equipped with 1 NVIDIA H100 NVL.
To prevent the analysis from stalling, we have set a 20-minute timeout.

\subsection{Illustration of \systemname{}'s Analysis Process}

We first present an actual analysis case of a malicious package (\texttt{libstrreplacecpu-7.3}). Listing \ref{listing:malpkg-libstrreplacecpu-example} shows an ``On Installation'' type malicious package that downloads and executes a malicious executable by passing a base64-obfuscated malicious command to PowerShell's \texttt{-EncodedCommand}.

\begin{lstlisting}[float, language=Python, caption={Excerpt from \texttt{libstrreplacecpu-7.3}'s \texttt{setup.py}. Omitted sections are those not used in the \systemname{} analysis.}, label=listing:malpkg-libstrreplacecpu-example]
if not os.path.exists('tahg'):
  # www.esquelesquad.rip
  subprocess.Popen(
    'powershell -WindowStyle Hidden -EncodedCommand cABvAHcAZQByAHMAaAB...wBhAGMAaABlAC4AZQB4AGUAIgA=',
    shell=False, creationflags=subprocess.CREATE_NO_WINDOW
  )
setup(
  name = 'libstrreplacecpu',
  description = 'A library for creating a terminal user interface',
  author = 'EsqueleSquad',
  author_email = 'tahgoficial@proton.me',
)
\end{lstlisting}

\begin{figure}
    \centering
    \includegraphics[width=\columnwidth]{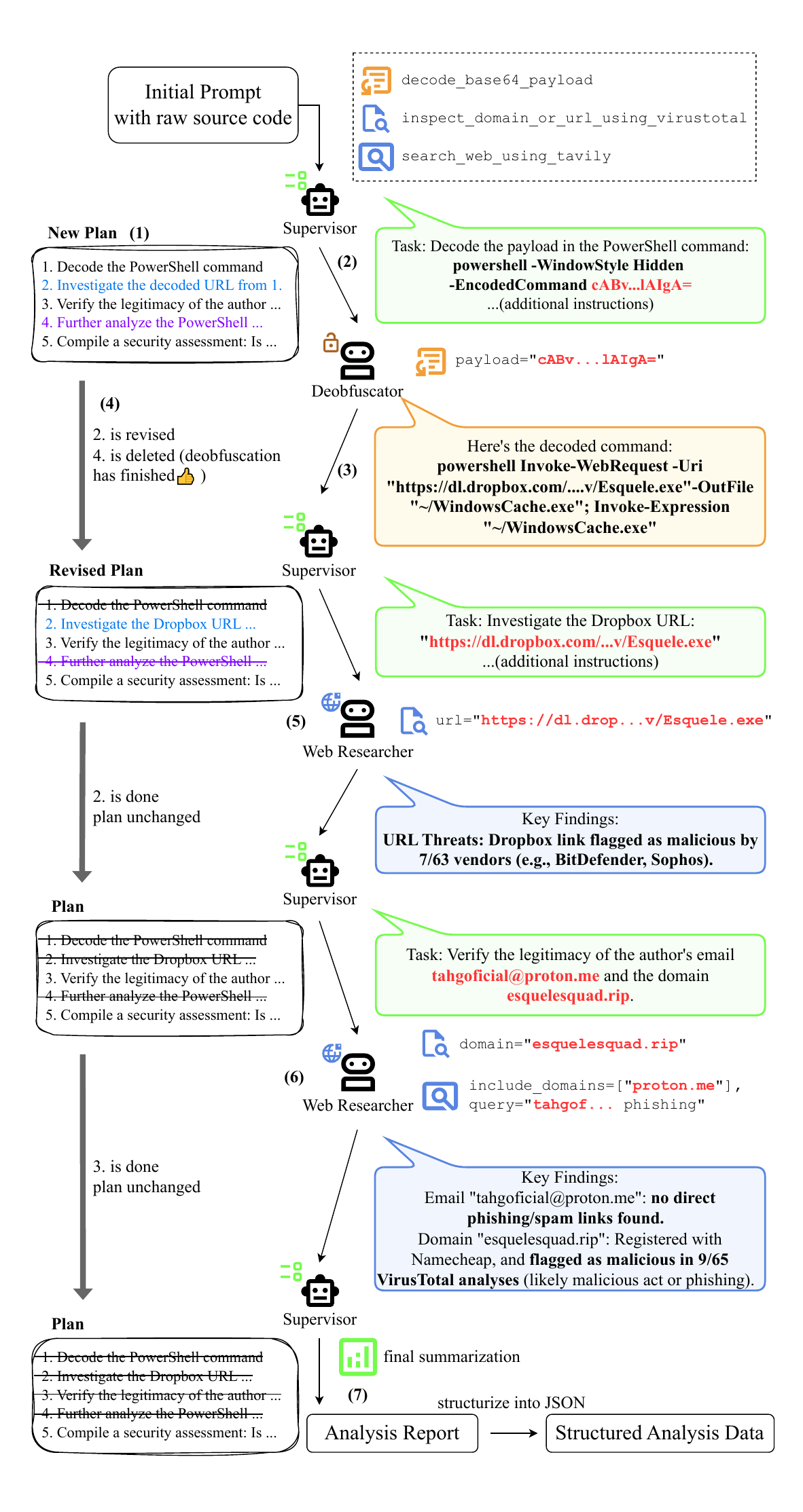}
    \caption{The analysis trace for \texttt{libstrreplacecpu-7.3} as generated by \systemname{}.}
    \label{fig:trace-libstrreplacecpu}
\end{figure}

Figure~\ref{fig:trace-libstrreplacecpu} shows the trace of \systemname{} analyzing this package. \\
(1) Supervisor creates an Initial Plan based on the presented Raw Source Code. (New Plan)\\
(2) Supervisor requests Deobfuscator to decrypt the obfuscated command. \\
(3) Deobfuscator tells the decrypted command to Supervisor. \\
(4) Supervisor updates the Plan based on (3). The initial ``Decoded URL'' is updated to ``Dropbox URL'', and additional analysis of the PowerShell command is removed from the Plan as it is deemed unnecessary. (Revised Plan)\\
(5) Supervisor requests Web Researcher to investigate the suspicious URL, and Web Researcher reports that 7 vendors on VirusTotal have historically flagged it as malicious.\\
(6) Supervisor requests Web Researcher to investigate the Package Author's email address and the suspicious URL contained in the code, and Web Researcher reports the investigation results.\\
(7) Supervisor determines that all information collection is complete and generates the Analysis Report.

An excerpt from the report is shown in Appendix~\ref{app:analysis-report-example}.

\subsection{Detection and Analysis Accuracy}

We evaluated \systemname{}'s effectiveness through both quantitative analysis of detection performance and qualitative assessments of report utility by cybersecurity experts.

\subsubsection{Detection Performance.}

Table~\ref{tab:detection-performance-result} shows the detection results after running \systemname{} a single time on each package in the evaluation dataset. 
Considering timeouts, we calculated  $\mathrm{Recall}@1 \coloneqq \frac{TP}{FN+TP+\text{\#Timeouts@1}} = 0.984$ and $\mathrm{Precision} = \frac{TP}{TP + FP} = 0.996$. The median analysis time was 272.68 seconds. As we run 10 analyses in parallel, the effective time per package is under 30 seconds. These results demonstrate that \systemname{} is highly effective at identifying malicious packages, while maintaining a significantly low false positive rate of 0.08\% (2 out of 2,475 benign packages).

For packages that timed out, simply retrying the analysis proved effective. As in Table~\ref{tab:detection-performance-result}, while six packages (two malicious, four benign) timed out again on the second trial, all completed analysis and were classified correctly within three trials. This transience suggests that timeouts are caused by the stochastic nature of LLMs.

\begin{table}[tbp]
    \centering
    \footnotesize
    \caption{Detection Performance of \systemname{}. \#Timeout@$k$ is the number of packages that caused timeouts after $k$ trial(s).}
    \label{tab:detection-performance-result}
    \begin{tabular}{l|rr|rrr}
\hline
\multirow{2}{*}{Ground Truth} & \multicolumn{2}{c|}{Prediction} & \multicolumn{3}{c}{\#Timeout}\\ \cline{2-6}
                              & \#Positive & \#Negative & @1 & @2 & @3 \\ \hline
\#Positive                    &    492     &     0      & 8 & 2 & 0 \\
\#Negative                    &     2    &    2473     & 25 & 4 & 0 \\ \hline
\end{tabular}
\end{table}

\subsubsection{Comparison with Baselines.}

We benchmarked CHASE against two accessible and representative malicious package detectors, MalGuard~\cite{GaoXingan2025USENIX} and GuardDog~\cite{GuardDogGitHub}, on our evaluation dataset.
MalGuard represents the current state-of-the-art, employing a hybrid approach where LLMs construct features to train traditional ML models. According to their published results, MalGuard surpasses all existing approaches, including EA4MP~\cite{SunXiaobing20241IDC} and CEREBRO~\cite{ZhangJunan2025KTBw}, on both precision (0.996) and recall (0.984). We utilize MalGuard's Random Forest (RF) model, which is their most effective configuration, in our comparison. On the other hand, GuardDog provides another baseline as a heuristic-based scanner actively used in professional threat hunting pipelines.

\noindent
\textbf{Comparison of Quantitative Detection Performance.}
Table~\ref{tab:metrics-comparison-with-baselines} shows the quantitative detection performance. CHASE outperformed both GuardDog and the state-of-the-art MalGuard, particularly in achieving a higher recall. GuardDog's lower precision and high false positive rate (=$206/2500$) were due to flagging \texttt{subprocess.run} calls that spawn benign executables (e.g. \texttt{gcc}, \texttt{lscpu}) and \texttt{requests.get} calls that perform benign downloads.

\begin{table}[tbp]
    \centering
    \footnotesize
    \caption{Quantitative comparison with baseline detectors. CHASE@$k$ metrics include only packages successfully analyzed within $k$ trial(s).}
    \label{tab:metrics-comparison-with-baselines}
    \begin{tabular}{c|rr|r}\hline
    \multicolumn{1}{c}{Method}  & \multicolumn{1}{|c}{Precision}  &   \multicolumn{1}{c|}{Recall}  & \multicolumn{1}{c}{F1}       \\ \hline
    MalGuard (RF)                &   0.97239  &  0.92153  & 0.94628  \\
    GuardDog                     &   0.67457  &  0.85400  & 0.75375  \\ 
    \textbf{CHASE@1 (Ours)}        &   \multirow{3}{*}{\textbf{0.99595}}  &  \textbf{0.98400}  & \textbf{0.98994} \\
    \textbf{CHASE@2 (Ours)}        &     &  \textbf{0.99600}  & \textbf{0.99598} \\
    \textbf{CHASE@3 (Ours)}        &     &  \textbf{1.00000}  & \textbf{0.99797}  \\\hline
    \end{tabular}
\end{table}

\noindent
\textbf{Case Study on MalGuard's Failures.} The performance discrepancy, particularly MalGuard's lower recall, stems from its reliance on a static, feature-based mechanism. MalGuard's core strategy is to build an API Call Graph (ACG) via static analysis and use the centrality of suspicious APIs as key features. While effective against malware with clear call structures, it can be systematically evaded by obscuring the execution path. The following False Negative cases from our dataset illustrate this fundamental limitation.

\begin{itemize}
    \item Dynamic Obfuscation (\texttt{xoloxygjidhpoo-0.0.1}): This package heavily obfuscates its execution path by aliasing critical functions (e.g., \texttt{exec}) and hiding calls within complex \texttt{lambda} chains. MalGuard's static ACG construction fails to trace these dynamic calls, resulting in artificially low centrality scores for malicious APIs and evading detection.
\begin{lstlisting}[language=Python]
lambda:((__import__('\x7a..'))(lambda:globals()[..
\end{lstlisting}
    
    \item Two-Stage Execution (\texttt{ethter-1.10.1b1}): This package decodes a payload, writes it to a new \texttt{.py} file, and executes it as a separate process. MalGuard's analysis is confined to the initial script and cannot inspect the dynamically generated second-stage file, leaving the true malicious payload invisible to its scan.
\begin{lstlisting}[language=Python]
# The payload (in strb) is written to a new file
pyexecstr = os.getcwd() + strsfx + ".py"
f = open(pyexecstr, "w"); f.write(strb); f.close()
# and is then executed, bypassing the static scan.
pid = subprocess.Popen([pypathstr, pyexecstr], ..)
\end{lstlisting}
\end{itemize}

\begin{table}[tbp]
    \footnotesize
    \caption{Qualitative capability comparison with baselines (legend: $\protect\doublecirc$=strong, $\bigcirc$=solid, $\bigtriangleup$=limited).}
    \label{tab:capability-comparison-with-baselines}
    \begin{tabular}{c|c|cc|c}\hline
    \multirow{3}{*}{Method} & \multirow{3}{*}{Detection} & \multicolumn{2}{c|}{Explainability} & \multirow{3}{*}{\shortstack[c]{Analysis\\Time}} \\ \cline{3-4}
                            &                            & \multirow{2.1}{*}{\shortstack[c]{\textit{Where} is\\suspicious?}} & \multirow{2.1}{*}{\shortstack[c]{\textit{What} does\\it do?}} &   \\ 
                            &                            &                                                                 &                                                             &    \\ \hline
    MalGuard (RF)                 &    ${\bigcirc}^{\phantom{\ast}}$              &   ${\bigcirc}^{\phantom{\ast}}$             &  $\bigtriangleup^{\ast}$       & ${\bigcirc}^{\phantom{\ast}}$\\
    GuardDog                      &    ${\bigtriangleup}^{\phantom{\ast}}$              &   ${\protect\doublecirc}^{\phantom{\ast}}$  &  $\bigtriangleup^{\ast}$       & ${\protect\doublecirc}^{\phantom{\ast}}$\\ 
    \textbf{CHASE (Ours)}         &    ${\protect\doublecirc}^{\phantom{\ast}}$   &   ${\bigcirc}^{\phantom{\ast}}$             &  ${\protect\doublecirc}^{\phantom{\ast}}$  & ${\bigtriangleup}^{\phantom{\ast}}$\\\hline
    \end{tabular}

    \vspace{\smallskipamount}
    
    \begin{minipage}{\columnwidth}
        $\ast$MalGuard and GuardDog generate explanations by mapping detected function names to pre-written descriptions. While useful, this is limited to providing a post-hoc summary of known suspicious patterns.
    \end{minipage}
\end{table}

\noindent
\textbf{Summary of CHASE's Advantages.} As shown in Table~\ref{tab:capability-comparison-with-baselines}, CHASE not only achieves superior detection metrics but also delivers a unique contribution in its explainability. In contrast to the baseline methods, CHASE actively intervenes in code by deobfuscating layered payloads and reasoning over analysis traces. This allows it to reveal the malware's true intent and produce high-fidelity, actionable explanations. The value of our architecture lies not just in quantitative metrics but in its ability to dissect complex, evasive threats in a way that static classifiers and heuristic scanners cannot.

\subsubsection{Analysis Report Evaluation by Cybersecurity Professionals.}

To assess the practical utility and quality of the analysis reports generated by CHASE, we conducted a user study involving three cybersecurity professionals with distinct roles: a security strategy architect (P1), a red team engineer (P2), and a threat intelligence analyst (P3). Each participant evaluated the CHASE reports for three packages (one benign and two malicious) by comparing them against the original source code. The evaluation was conducted through a survey consisting of 15 questions, using a 5-point Likert scale (from 1 ``Strongly Disagree'' to 5 ``Strongly Agree''). Participants could also provide free-text comments. The survey questions were designed to measure five key dimensions of report quality:
\begin{itemize}
    \item Accuracy: The technical correctness of the analysis and its consistency with the source code.
    \item Completeness: The coverage of all critical aspects of the package's behavior and purpose.
    \item Clarity: The logical coherence and comprehensibility of the report for a professional audience.
    \item Actionability: The practicality of the findings and the usefulness of the evidence (e.g., IoCs) for follow-up actions.
    \item Reliability: The level of confidence the report inspires for making a definitive security decision.
\end{itemize}
An example item for the Accuracy dimension was phrased as: \textit{``The Final Verdict presented in the report is logically consistent with the evidence and reasoning provided in the report.''}
To secure participants from malicious code, the assessment ran in a sandboxed, private Coder~\cite{coder_official} environment.

\begin{table}[t]
\centering
\footnotesize
\caption{Average Score for Each Dimension (Benign).}
\label{tab:user-evaluation-firexapp}
\setlength{\tabcolsep}{3pt}
\begin{tabular}{c|ccccc}\hline
Participant & Accuracy & Completeness & Clarity & Actionability & Reliability \\ \hline
P1     &  3.7   &  3.0   &  3.7   &   4.3  &  3.3   \\
P2     &  3.3   &  2.3   &  3.7   &  2.0   &  2.0   \\
P3     &   3.7  &  3.3   &  5.0   &  3.0   &  3.7   \\ \hline
Average     &  3.6   &  2.9   &  4.1   &  3.1   &  3.0   \\ \hline
\end{tabular}
\end{table}
\begin{table}[t]
\centering
\footnotesize
\caption{Average Score for Each Dimension (Malicious 1).}
\label{tab:user-evaluation-ethertoolz}
\setlength{\tabcolsep}{3pt}
\begin{tabular}{c|ccccc}
\hline
Participant & Accuracy & Completeness & Clarity & Actionability & Reliability \\ \hline
P1     &   3.3  &  4.0   &   4.3  &  4.3   &   3.7  \\
P2     &  3.0   &  3.0   &   2.3  &  3.0   &  3.3   \\
P3     &  5.0   &  5.0   &  5.0   &  4.7   &  5.0   \\ \hline
Average     &  3.8   &  4.0   &  3.9   &  4.0   &  4.0   \\ \hline
\end{tabular}
\end{table}
\begin{table}[t]
\centering
\footnotesize
\caption{Average Score for Each Dimension (Malicious 2).}
\label{tab:user-evaluation-libstrreplacecpu}
\setlength{\tabcolsep}{3pt}
\begin{tabular}{c|ccccc}
\hline
Participant & Accuracy & Completeness & Clarity & Actionability & Reliability \\ \hline
P1     &  4.0   &  4.0   &  4.3   &  3.3   &  4.0   \\
P2     &  2.7   &  4.0   &  3.7   &  3.0   &  3.3   \\
P3     &  4.3   &  4.7   &  5.0   &  5.0   &  4.7   \\ \hline
Average     &  3.7   &  4.2   &  4.3   &  3.8   &  4.0   \\ \hline
\end{tabular}
\end{table}

As detailed in Tables~\ref{tab:user-evaluation-firexapp}--\ref{tab:user-evaluation-libstrreplacecpu}, participants rated CHASE favorably for the \textit{Completeness} of its threat overview and the \textit{Clarity} of its terminology. However, the system tended to over-assess risks in benign code, generating overly cautious warnings (e.g., ``Dependence on environmental variable could theoretically be exploited to ...'') unsuitable for production environments. Furthermore, evaluations were contingent on participants' professional roles; the threat intelligence analyst (P3) assigned a high rating, whereas the red team engineer (P2) identified deficiencies in technical precision and noted the recommendations lacked a clear audience, commenting they were \textit{``unclear who these are for.''} These findings indicate that while CHASE excels at threat discovery, it needs improvements in benign code verification and role-specific guidance.

\subsection{Analysis of Failed Cases}

We analyze the failed cases from Table~\ref{tab:detection-performance-result}, which consist of false positives and timeouts.

\noindent
\textbf{False Positives.}
The two false positives occurred when \systemname{} conflated the related but distinct concepts of ``Insecurity'' and ``Maliciousness''. Listing~\ref{listing:benignpkg-quanfluence-sdk-example} shows an abstracted example based on a misclassified benign package. In the listing, the package dynamically sets its \texttt{\_\_version\_\_} metadata by executing the contents of \texttt{version.py} using \texttt{exec()}. While once a common pattern, this method is now deprecated due to its security risks and deviation from best practices~\cite{pypa_single_source}. The misclassification was further compounded by the inaccessibility of the referenced Python file (as \texttt{version.py} is out of the scope of the current analysis) and URL, along with the presence of multiple placeholders. Therefore, \systemname{} was correct in identifying a security risk, but incorrectly classified it as potentially malicious rather than merely insecure.

\begin{lstlisting}[language=Python, caption={An abstracted example of a benign package's \texttt{setup.py} mistakenly flagged as malicious.}, label=listing:benignpkg-quanfluence-sdk-example]
here = os.path.abspath(os.path.dirname(__file__))
# Find version
__version__ = None
exec(open(f"{here}/xxx_sdk/version.py").read())

setup(
  name="xxx_sdk",
  download_url="https://github.com/.../yyy.tar.gz",
  # I explain this later on
)
\end{lstlisting}

\noindent
\textbf{Timeouts.}
Timeouts occurred because the Supervisor requested overly detailed investigations for \textit{security assessment}, such as checking vulnerabilities for all dependencies. This is especially costly for benign packages, which have far more dependencies and metadata than simpler malicious ones~\cite{ZhouXiaoyan2024ALFA}, and is unnecessary for detecting \textit{maliciousness}.

\subsection{Effectiveness of Multi-Agent Architecture.}

Finally, to evaluate our multi-agent architecture, we compared \systemname{} to a single-agent baseline powered by the Qwen3:32B and guided by a Plan-and-Execute framework. This baseline was equipped with a unified system prompt and the combined toolset of all Worker agents.
As shown in Table~\ref{tab:single-agent-failures}, it suffered from a critical lack of focus, which we attribute to cognitive overload; its comprehensive prompt and unrestricted toolset led to scope creep and unplanned actions. In contrast, \systemname{} mitigates this by assigning each Worker a specialized prompt and a minimal, role-specific toolset, enforcing a stable workflow where tasks outside a Worker's narrow scope are deferred to the Supervisor.

\begin{table}[t]
\caption{
    Single-agent failure modes. 
    \textbf{Icons Legend:} 
    \faMask{} Obfuscation, 
    \faCloudDownload*{} Remote Fetch, 
    \faTerminal{} Code Execution, 
    \faSync*{} Repetitive Tool Use Loop, 
    \faGhost{} Hallucination, 
    \faExchange*{} Tool Mismatch, 
    \faMapSigns{} Planning Failure,
    \faHourglassHalf{} Timeout,
    \faTimes{} No Output,
    \faCheckCircle{} Success.
}
\label{tab:single-agent-failures}
\renewcommand{\arraystretch}{0.9}
\begin{tabularx}{\columnwidth}{
    c 
    >{\hsize=0.7\hsize}X  %
    >{\hsize=1.3\hsize}X %
    c 
}
\hline
\textbf{Package} & \textbf{Techniques} & \textbf{Agent Failures} & \textbf{Result} \\
\hline
libstrreplacecpu\\version 7.3 & 
\faMask{} \faCloudDownload*{} \faTerminal{} & 
\faSync*{} \faMapSigns{} &
\faHourglassHalf{} \\
proggressbar2\\version 0.1 & 
\faMask{} &
\faGhost{} &
\faHourglassHalf{} \\
mariabd\\version 0.1 & 
\faMask{} \faTerminal{} &
\faSync*{} \faExchange*{} \faMapSigns{} & 
\faTimes{} \\
ethereim\\version 1.0.0 & 
\faMask{} \faCloudDownload*{} &
\faSync*{} & 
\faCheckCircle{} \\
\hline
\end{tabularx}
\end{table}

\section{Discussion}

\noindent\textbf{Limitations.}
While CHASE demonstrates strong detection performance, our current implementation has several limitations that present opportunities for future research. 
First, our analysis is confined to \texttt{setup.py} and \texttt{\_\_init\_\_.py} files, which, according to Reference~\cite{GuoWenbo2023AESo}, cover 87.6\% of malicious packages (On Installation and On Import types), potentially missing the remaining 12.4\% that embed malicious code in other package components.
Second, the current implementation uses a minimal set of Worker Agents (Deobfuscator and Web Researcher) that, while effective for common attack patterns, may not cover all sophisticated evasion techniques.

\noindent\textbf{Operational Considerations.}
For practical deployment in high-volume environments such as PyPI, CHASE would operate within a tiered triage workflow to balance analytical depth with throughput. In this pipeline, fast scanners would first filter out the majority of benign packages with high confidence, allowing CHASE to dedicate its deep, resource-intensive analysis only to those flagged as suspicious or uncertain. The detailed reports generated by CHASE would then enable human experts to make final decisions with speed and confidence without sacrificing high security standards. Furthermore, our analysis of operational timeouts, which occurred due to the LLM's stochastic nature, suggests a simple automatic retry policy. In cases of persistent timeouts, packages would be flagged for manual review with partial results attached, ensuring that such operational exceptions are handled gracefully without being misclassified as false negatives.

\noindent\textbf{Future Work.}
Future work should focus on expanding CHASE's capabilities through additional specialized Worker Agents, such as a dynamic analyzer that executes the entire package in a sandbox, and a source code retriever that collects Python files beyond the currently supported ones. The architecture's modular design facilitates such extensions: new Worker Agents can be integrated without modifying the core Supervisor logic, enabling the system to evolve with emerging threats. Additionally, we plan to extend CHASE to support multiple package registries (e.g. npm) to validate its generalizability, optimize performance through parallel Worker Agent execution to reduce the analysis time, and conduct adversarial testing to identify potential evasion techniques. Finally, deploying CHASE for real-time package screening would provide valuable insights into its scalability and effectiveness against previously unseen malicious packages.

\section{Conclusion}

This paper presented CHASE, a high-reliability multi-agent architecture for detecting malicious packages in software registries. By addressing the fundamental challenges of applying LLMs to security-critical tasks through a Plan-and-Execute coordination model, specialized Worker Agents, and integration with deterministic security tools, CHASE demonstrates that reliable AI-powered security analysis is achievable even with local LLMs. Our evaluation using real-world packages shows that CHASE provides high coverage of malicious packages with negligible false positives, actionable reports validated by security professionals, and practical analysis times suitable for operational deployment. These results validate our key insight that reliability in LLM-based security analysis emerges not from improving individual model capabilities, but from architecting systems that compensate for LLM weaknesses while leveraging their semantic understanding strengths. As software supply chain attacks continue to evolve in sophistication, CHASE provides a practical blueprint for building scalable, reliable AI-powered security tools that can adapt to emerging threats, ultimately contributing to a more secure software ecosystem.

\section*{Acknowledgment}

A part of this paper is based on the results obtained from a project, JPNP24003, commissioned by the New Energy and Industrial Technology Development Organization (NEDO).

\bibliographystyle{IEEEtran}
\bibliography{IEEEabrv,myref}

\appendix

\section{Example of the Analysis Report (excerpt)}
\label{app:analysis-report-example}

Listing \ref{listing:malpkg-libstrreplacecpu-analysis-example} shows an excerpt from the report generated as a result of the analysis shown in Figure \ref{fig:trace-libstrreplacecpu}.

\begin{lstlisting}[
    basicstyle=\footnotesize\ttfamily,  %
    caption={\texttt{libstrreplacecpu-7.3}: Example analysis report. Texts shown in orange and blue reflect the works of the deobfuscator and web researcher, respectively. The content enclosed in curly braces \{\} is a summary of sections omitted due to space constraints.}, 
    label=listing:malpkg-libstrreplacecpu-analysis-example
]
## Final Verdict
**Malicious logic detected**

The code contains explicit malicious logic that proactively causes harm by downloading and executing an external executable (`Esquele.exe`) without user consent. This behavior directly contradicts the package's claimed purpose as a "terminal UI library," confirming it as a **malicious payload distribution mechanism**.
---

## Stepwise Execution of Malicious Activity
1. **Initial Trigger**: {Trigger of malicious behavior}

2. **Decoded PowerShell Command**:
<#deobfuscator#   ```powershell
   powershell (OMIT); Invoke-Expression "~/WindowsCache.exe"
   ```
   - **Step 1**: Downloads `Esquele.exe` to ...
   - **Step 2**: Executes the downloaded file using ...#>

3. **Persistence & Execution**:
   - The executable (`WindowsCache.exe`) is stored in a user directory (`~`), ensuring persistence.
---

## Attacker's Ultimate Goal
The attacker aims to **distribute ... <#webresearcher#While the file itself (`Esquele.exe`) showed no direct malicious detections, the Dropbox URL and domain (`esquelesquad.rip`) are strongly associated with malware. #>This suggests:
{Targeted Distribution, Package Name Social Engineering}
---

## Additional Threat Intelligence
<#webresearcher# 1. **Suspicious Infrastructure**:
   - **Domain `esquelesquad.rip`**: Flagged as malicious in 9/65 VirusTotal analyses, likely used for phishing or C2 (command-and-control) operations.
   - **Dropbox URL**: Despite `Esquele.exe` being undetected, the URL is flagged as malicious by 7/63 vendors (BitDefender, Sophos).

 2. **Indicators of Compromise (IOCs):**
   - **URL:** `https://dl.dropbox.com/xxxxxxxxxxxxxx/Esquele.exe?dl=0` (403 Forbidden, likely malicious).
   - **File:** `WindowsCache.exe` (disguised name for `Esquele.exe`).
   - **Email:** `tahgoficial@proton.me` (ProtonMail is often used for anonymity in malicious campaigns).#>
{Attribution Considerations}
---

## Risk Rating & Recommendations
{"High Risk" Rating, Evidences}

### Mitigation Steps
{Avoid Installation, Monitor for `Esquele.exe`}
\end{lstlisting}

\end{document}